\documentclass[journal]{IEEEtran}
\usepackage{graphicx}
\usepackage{subfigure}
\usepackage{float}
\usepackage{amsmath}
\usepackage{epstopdf}
\usepackage{cases}
\usepackage{color}
\usepackage{algorithm}
\usepackage{algpseudocode}
\usepackage{amsmath}
\usepackage{amssymb}
\usepackage{caption}
\makeatletter
\renewcommand{\maketag@@@}[1]{\hbox{\m@th\normalsize\normalfont#1}}%
\makeatother
\usepackage{stfloats}
\usepackage{cite}
\usepackage{makecell}
\usepackage{multirow}

\ifCLASSINFOpdf
\else
\fi

\usepackage{caption}
\usepackage{mathtools}

\begin{document}

\title{Integrated Sensing and Communication \\Channel Modeling: A Survey}
	
\author{Zhiqing~Wei,~\IEEEmembership{Member,~IEEE,}
	Jinzhu~Jia,
	Yangyang~Niu,~\IEEEmembership{Student Member,~IEEE,}
	Lin~Wang,~\IEEEmembership{Student Member,~IEEE,}
	Huici~Wu,~\IEEEmembership{Member,~IEEE,}
	Heng~Yang,	
	Zhiyong~Feng,~\IEEEmembership{Senior Member,~IEEE}
	
	\thanks{Zhiqing~Wei, Jinzhu~Jia, Yangyang~Niu, Lin~Wang, Heng~Yang and 
		Zhiyong~Feng are with Key Laboratory of Universal Wireless Communications, Ministry of Education, Beijing University of Posts and Telecommunications (BUPT), Beijing 100876, China (emails: weizhiqing@bupt.edu.cn, jiajinzhu@bupt.edu.cn, niuyy@bupt.edu.cn, wlwl@bupt.edu.cn, dailywu@bupt.edu.cn, yangheng@bupt.edu.cn, fengzy@bupt.edu.cn).
	
	Huici~Wu is with the National Engineering Lab for Mobile Network
	Technologies, Beijing University of Posts and Telecommunications (BUPT), Beijing
	100876, China (e-mail: dailywu@bupt.edu.cn).}}

\maketitle

\begin{abstract}
Integrated sensing and communication (ISAC) is expected to play 
a crucial role in the sixth-generation (6G) mobile communication systems, 
offering potential applications in the scenarios of intelligent transportation, 
smart factories, etc. 
The performance of radar sensing in ISAC systems is closely 
related to the characteristics of radar sensing and communication channels. 
Therefore, ISAC channel modeling serves as a fundamental cornerstone 
for evaluating and optimizing ISAC systems. 
This article provides a comprehensive survey 
on the ISAC channel modeling methods. 
Furthermore, the methods of target radar cross section (RCS) modeling 
and clutter RCS modeling are summarized. 
Finally, we discuss the future research trends related to 
ISAC channel modeling in various scenarios.
\end{abstract}


\begin{figure*}[!htbp]
\begin{center}
\includegraphics[scale=1.2]{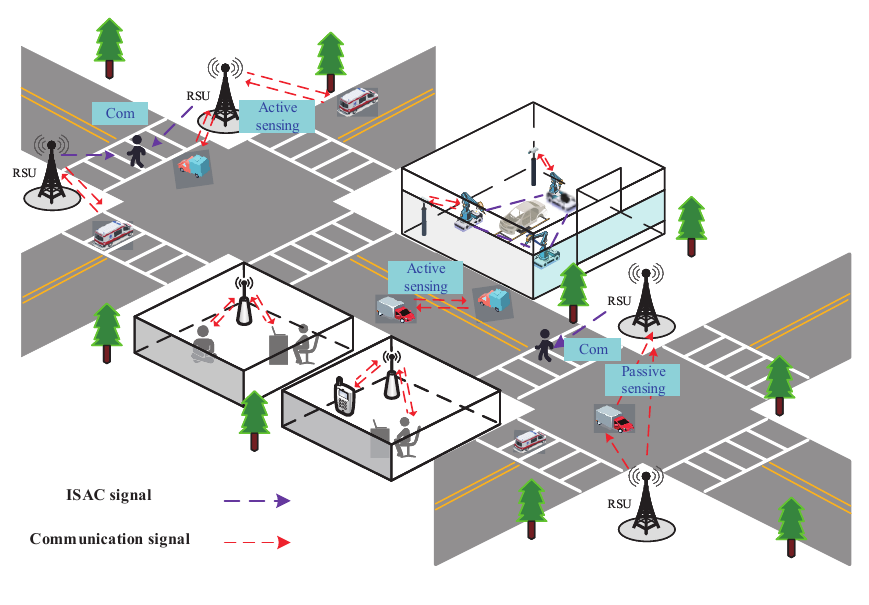}
\end{center}
\caption{Application scenarios of ISAC systems. }
\label{ISAC}
\end{figure*}

\begin{table*}[!htbp]
\caption{List of abbreviations}
\label{tab_g}
\renewcommand{\arraystretch}{1.5} 
\begin{center}
	\begin{tabular}{m{0.1\textwidth} m{0.3\textwidth} m{0.1\textwidth} m{0.3\textwidth}}
		\hline			
		3GPP & 3rd Generation Partnership Project &
		6G & Sixth-generation \\
            BS & Base station &
		PDF & Probability density function \\		
		CDL & Clustered delay line &
		DoF & Degree of freedom \\
		EM & Electromagnetic methods &
		FDTD & Finite-difference time-domain \\
		FMM & Fast multipole method &
		GC-PDF & Generalized compound probability density function \\
		GMDM & Gaussian mixture density model &
		GO & Geometric optics \\
		ISAC & Integrated sensing and communication &
		LOP & Legendre orthogonal polynomials \\
		LoS & Line-of-sight &
		MoM & Method of moment \\
		NLoS & Non-line-of-sight &
		PO & Physical optics \\
		RCS & Radar cross section &
		SBR & Signal-based ray-tracing \\
		SPM & Stationary phase method &
		TDL & Tapped delay line \\
        UE  & User equipment &
		UTD & Uniform theory of diffraction  \\		
		\hline
	\end{tabular}
\end{center}
\end{table*}
	
\section{Introduction}

Integrated sensing and communication (ISAC) 
provides the dual functions of radar sensing and communication in the same system, 
which is of great significance in the scenarios such as intelligent transportation, smart factories \cite{cui2021integrating,wei2024integrated}, as shown in Fig. \ref{ISAC}. 
ISAC achieves high-rate and low-latency communication, 
as well as high sensing accuracy by exploiting 
the mutual benefit between radar sensing and communication \cite{feng2020joint,liu2022survey}. 
The future mobile communication systems are characterized by 
high frequency bands, large bandwidth, and large antenna array, 
which are beneficial to sensing, 
enabling the implementation of ISAC technology. 

The radar channel model reveals the characteristics of radar signal propagation, 
which is the cornerstone for signal processing, interference management, 
and performance evaluation of ISAC system \cite{jiang2023integrated}. 
In the 3rd Generation Partnership Project (3GPP) standards, 
random channels are usually applied to evaluate the performance of communication 
due to the low computational complexity. 
However, it lacks the modeling of the radar scattering cross sectional 
product of targets and environmental scatterers, which is essential for radar sensing. 
Fishler \emph{et al.} \cite{fishler2004mimo} first proposed the concept of 
``radar channel'' considering the radar cross section (RCS) characteristic of target, 
which is mainly modeled by deterministic modeling methods \cite{herman2004joint} and 
statistical modeling methods \cite{weinstock1964target}. 
The deterministic modeling methods have 
high computational complexity and limited application scenarios. 
Although the statistical modeling methods have a wide range of applications, 
they rely on a large amount of channel measurements. 

The deterministic modeling methods utilize the principles of 
electromagnetic wave propagation to accurately predict the radar channel model. 
The accurate methods refer to solving the 
electromagnetic wave equation or the integral equation of 
the induction field distributed over the surface of target, 
including geometric optics (GO) \cite{chatzigeorgiadis2004development, perez1994application}, 
physical optics (PO) \cite{uluisik2008radar}, 
signal-based ray-tracing (SBR) \cite{tao2008kd, bennani2012rcs}, etc. 
There have been some studies on the deterministic modeling of RCS. 
Adana \emph{et al.} \cite{1364146} proposed an RCS area calculation method 
for mono-static radar with complex targets based on the PO method and 
the stationary phase method (SPM). 
Zhang et al. \cite{zhang2018multibeam} employed the ray-tracing method 
to predict radar signal propagation, achieving precise estimation of target RCS.
As for statistical modeling methods, 
Swerling \emph{et al.} \cite{swerling1997radar, swerling1970recent} 
proposed five statistical models for the fluctuation characteristic of targets in the 1960s, 
namely Swerling I to V models. 
In \cite{meyer1973radar}, several statistical models for the RCS of targets 
are introduced, including the chi-square model, 
the log-normal distribution model, and the Weibull model.

Moreover, clutter modeling also plays an essential role in radar channel modeling. 
Similar to the RCS modeling of target, 
the deterministic modeling and statistical modeling methods can be applied in clutter modeling. 
The RCS characteristics of clutter sources are 
closely related to the scenarios and the parameters of radar sensing. 
Deterministic modeling of clutter predicts the RCS characteristics of clutter cells. 
Taking the sea clutter as an example, 
the well-known deterministic models include the Georgia Institute of Technology (GIT) model \cite{paulus1990evaporation}, 
hybrid model \cite{dockery1990method}, and Morchin model \cite{rountree1990radar}.
Furthermore, the clutter can be characterized by statistical modeling 
due to the random fluctuation of the amplitude of clutter. 
Drosopoulos \emph{et al.}  \cite{drosopoulos1994description} 
verified that the sea clutter approximately follows the Gaussian distribution. 
The classical statistical models such as Marcum \cite{marcum1960statistical} and 
Swerling models \cite{swerling1960probability} are applicable to low-resolution radars. 
However, the above-mentioned classical models fail to provide accurate predictions for 
the channel modeling of high-resolution radars \cite{gerlach1999spatially}. 
Therefore, the log-normal, Weibull, and $K$-distribution models 
are proposed \cite{george1968detection}. 
Nevertheless, it is difficult to accurately describe the characteristics of 
clutter using a single statistical model \cite{ward2010use}. 
To accurately describe the temporal and spatial characteristics of clutter, 
the compound statistical modeling methods are proposed by adjusting the model parameters
\cite{middleton1999new}\cite{dong2006distribution}.

The clustered delay line (CDL) and tapped delay line (TDL) channel models are 
used to model the communication channels such as the one-way channel 
between the transmitter and the receiver in 3GPP TR 38.901 \cite{3gpp2018study}. 
However, in the ISAC systems, the transceiver detects the target 
through the received echo signal reflected from target. 
Thus, the radar channel needs to establish a round-trip channel model. 
The above-mentioned communication channels can be modified by 
introducing coefficients such as RCS to 
model the radar channel \cite{chen2021code}. 
Hence, it is feasible to establish radar channel models based on 
the communication models, which provide references for radar channel modeling. 
Overall, the differences between communication channels 
and radar channels are summarized as follows.

\begin{itemize}
	\item[$\bullet$] \textbf{Difference in the locations of transceiver:} The transmit communication data is inherently stochastic \cite{zhang2015wireless}. Consequently, the transmitter and receiver in communication system are spatially separated. The sensing signal is well-defined and structured to estimate the location, velocity, or image of target. Radar sensing modes consist of active sensing and passive sensing \cite{swerling1960studies}. In active sensing, the transmitter and receiver are co-located, while they are separately located in passive sensing. 
\end{itemize}
\begin{itemize}
	\item[$\bullet$] \textbf{Difference in the application scenarios:} Communication systems perform channel estimation based on received communication signal \cite{zhang2015wireless}. Conversely, radar system estimates the distance, location, direction, and image of target using the received echo signal \cite{swerling1960studies}. In communication system, multipaths are beneficial to spatial diversity. However, in radar system, active sensing relies on a line-of-sight (LoS) link for target detection and the parameters estimation, while a non-line-of-sight (NLoS) link will lead to false alarm \cite{wright2000space}. 
\end{itemize}
\begin{itemize}
	\item[$\bullet$] \textbf{Difference in channel fading:} Channel fading is mainly characterized by large-scale and small-scale fading in communication systems, modeling the propagation loss of communication signals over a transmission range. However, the fading coefficients are mainly related to the scattering characteristics of target in radar sensing channel, including target RCS and clutter RCS.
\end{itemize}

The deviation in random channel modeling methods will impact the sensing accuracy. 
In contrast, deterministic channel modeling methods accurately 
characterize radar channel. 
Li \textit{et al.} proposed the GO, PO, 
and Electromagnetic method (EM) to model radar channel \cite{li2021integrated}. 
Jing \textit{et al.} proposed a measurement-based statistical channel modeling approach, 
which is applied to scatter-based spatial channel models \cite{jing2023measurement}.
Overall, there are rich research achievements in the area of ISAC channel models.
However, there are rarely survey articles summarizing ISAC channel modeling.

\begin{figure}[!htbp]
\includegraphics[scale=0.9]{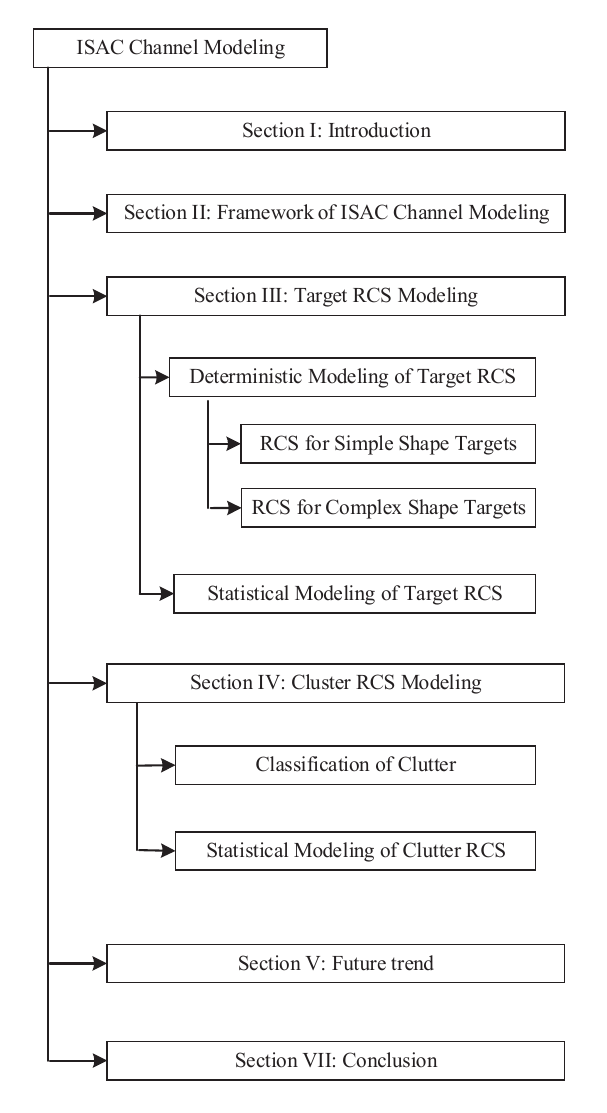}
\caption{The structure of this article.}
\label{The organization2}
\end{figure}

This article provides a comprehensive survey of the ISAC channel modeling. 
Specifically, the radar channel modeling and communication channel modeling methods are 
provided, including the modeling of target RCS and clutter RCS 
from deterministic and statistical perspectives. 
The structure of this article is illustrated in Fig. \ref{The organization2}. 
In Section \uppercase\expandafter{\romannumeral2}, 
we provide a brief overview of the channel modeling of ISAC 
in both active and passive sensing modes. 
Section \uppercase\expandafter{\romannumeral3} and \uppercase\expandafter{\romannumeral4} 
provide the modeling methods for target RCS and clutter RCS, respectively. 
Section \uppercase\expandafter{\romannumeral5} summarizes 
the future research trends and 
Section \uppercase\expandafter{\romannumeral6} concludes this article. 
The abbreviations used in this article are listed in Table \ref{tab_g}.

\section{Framework of ISAC Channel Modeling}\label{sec:system-model} 

In this section, the framework of ISAC channel modeling is introduced. 
Specifically, The communication and radar sensing channel modeling methods 
are reviewed for active and passive sensing, respectively.

\subsection{Active Sensing}

\begin{figure}[!htbp]
	\includegraphics[width=0.49\textwidth]{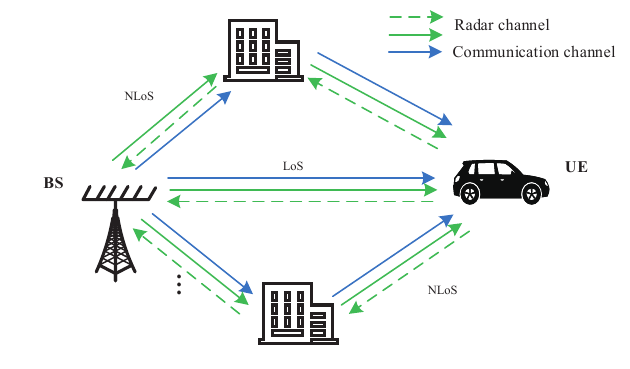}
	\caption{ISAC channel of active sensing.}
	\label{chch11}
\end{figure}

In active sensing mode, 
the transmitter and receiver are deployed in the same device. 
The active sensing device can be either the base station (BS) 
or the user equipment (UE). 
Take the BS as an example, the BS transmits ISAC signal to UE, 
while receiving the echo signal for target sensing, 
as shown in Fig. \ref{chch11}. 
The ISAC channel include the one-way communication channel 
following the path ``BS$\to$UE'' and 
the two-way sensing channel following the path ``BS$\to$target$\to$BS''.

\subsubsection{One-way communication channel}

The one-way communication channel can be expressed as\cite{chen2021code}

\begin{equation}
	\begin{array}{c}
		{\mathbf{H}_{com}^{act}} = \sqrt {{N_t}{N_r}} \underbrace {\tilde \alpha _0\mathbf{A}_r\left( {\theta _r^0} \right)\mathbf{A}_t^H\left( {\theta _t^0} \right)}_{\expandafter{(a)}}\\
		+ \sqrt {{N_t}{N_r}} \underbrace {\sum\limits_{l = 1}^{{L_{path}} - 1} {\tilde \alpha _l\mathbf{A}_r\left( {\theta _r^l} \right)\mathbf{A}_t^H\left( {\theta _t^l} \right)} }_{\expandafter{(b)}},
	\end{array}\label{hcomactive}
\end{equation}
where $(a)$ is the LoS path component and $(b)$ is the multipath component;
${N_t}$ and $ {N_r}$ are the numbers of the transmitting and receiving antenna array elements, respectively; 
$\tilde \alpha _l$ is the fading coefficient of the $l$-th propagation path; 
$ {{\mathbf{A}}_{r}}\left( {\theta _r^l} \right)$ 
and $ {\mathbf{A}}_{t}^{H}\left( {\theta _t^l} \right)$ are the steering vectors of the receiving and transmitting antenna arrays for the $l$-th propagation path, respectively; 
$\theta _t^l$ and $ \theta _r^l$ are the angle of arrival and angle of departure related to the $l$-th propagation path, respectively; 
${L_{path}}$ is the number of propagation paths, 
and $l=0$ represents the LoS path.

\subsubsection{Two-way sensing channel}
As shown in Fig. \ref{chch11}, the received echo signals reflected from the target experience a two-way sensing channel, which can be expressed as\cite{chen2021code}
\begin{align}
	{\mathbf{H}_{rad}^{act}} &= \sqrt {{N_t}{N_r}} \tilde \alpha _0\tilde \sigma _0 \mathbf{A}_r\left( {\theta _r^0} \right) \mathbf{A}_t^H\left( {\theta _t^0} \right)\notag \\
	&+ \sqrt {{N_t}{N_r}} \sum\limits_{l = 1}^{{L_{path}} - 1} {\tilde \alpha _l\tilde \sigma _l\mathbf{A}_r\left( {\theta _r^l} \right)\mathbf{A}_t^H\left( {\theta _t^l} \right)},
	\label{hhh}
\end{align}
where $\tilde \sigma _0$ is the RCS of target;
$\tilde \sigma _l$ is the RCS of clutter generating by the
scatter other than target, 
which are the key parameters in the two-way sensing channel.
In addition, the parameters 
$ {{\mathbf{A}}_{r}}\left( {\theta _r^l} \right)$, 
$ {\mathbf{A}}_{t}^{H}\left( {\theta _t^l} \right)$, 
$ {{\mathbf{A}}_{r}}\left( {\theta _r^l} \right)$, 
$ {\mathbf{A}}_{t}^{H}\left( {\theta _t^l} \right)$,
$\theta _t^l$, and $ \theta _r^l$ have the 
same meaning to these parameters in (\ref{hcomactive}).

\subsection{Passive Sensing}

\begin{figure}[!htbp]
	\includegraphics[width=0.48\textwidth]{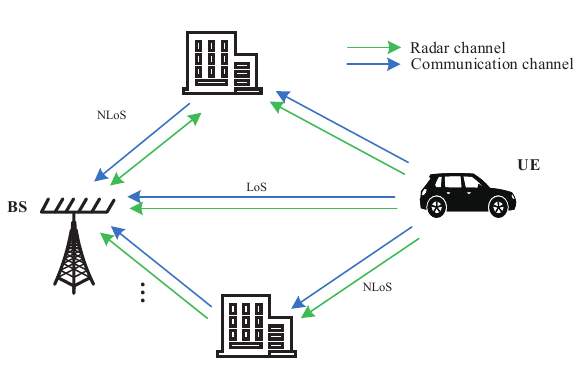}
	\caption{ISAC channel of passive sensing.}
	\label{chchch1}
\end{figure}

As illustrated in Fig. \ref{chchch1},
ISAC channel of passive sensing can be decomposed into the one-way communication channel and one-way sensing channel.
	
\subsubsection{One-way communication channel}

The one-way communication channel model in passive sensing 
is the same to that in active sensing.

\subsubsection{One-way sensing channel}

As shown in Fig. \ref{chchch1}, 
the radar signals in passive sensing experience 
one-way sensing channel from BS to UE.
However, the radar sensing channel of passive sensing 
has the same form to that of active sensing in (\ref{hhh}).
According to (\ref{hhh}),
the radar sensing channel models are closely related
with \textit{the RCS characteristics of target and scatter},
which will be reviewed in Sections \ref{Sec_Target_RCS}
and \ref{Sec_Clutter_RCS}, respectively.

\section{Target RCS Modeling}     
\label{Sec_Target_RCS}

The RCS is employed to characterize the signal 
attenuation caused by target, 
influenced by the target's geometric properties, 
surface material, and the wavelength of the incident electromagnetic 
wave \cite{levanon1988radar, cheng1989field}. 
The modeling methods for target RCS can be categorized into deterministic modeling and statistical modeling methods, which will be reviewed in this section.

\subsection{Deterministic Modeling of Target RCS}

\begin{figure}[!htbp]
	\includegraphics[scale=0.6]{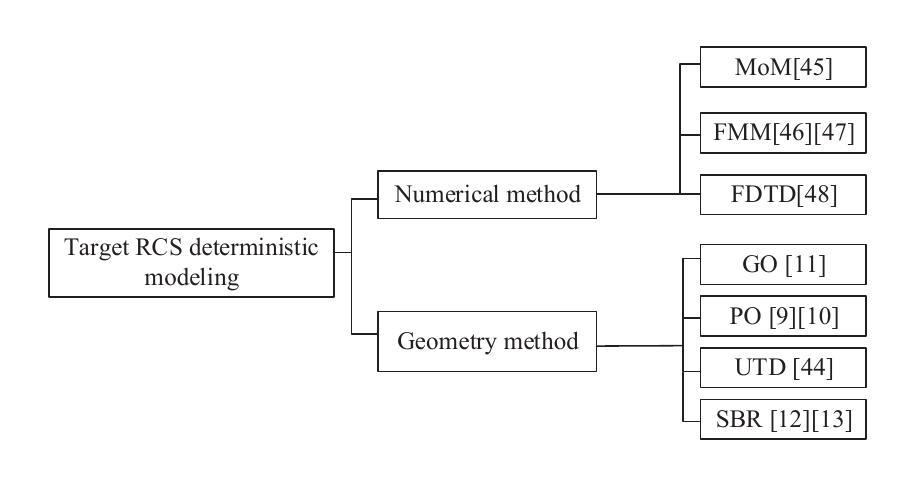}
	\caption{Deterministic modeling methods of target RCS \cite{herda2020radar}.}
	\label{A Classification}
\end{figure}

The deterministic modeling methods model 
the target RCS accurately and detailedly. 
Fig. \ref{A Classification} illustrates the classification of 
deterministic RCS modeling methods 
including geometry methods and numerical methods. 
Numerical methods calculate RCS accurately for 
the targets with complex shapes, 
which has high computational complexity and are better for small targets. 
In contrast, geometric methods provide fast calculation speed. 
However, they are only feasible for the target with 
size exceeding the wavelength.

\begin{table*}[htbp]
	\caption{ RCS of objects with several simple geometric shapes\cite{mahafza2005radar}}
	\renewcommand{\arraystretch}{1.7}
	\begin{center}
		\begin{tabular}{|m{0.3\textwidth}<{\centering}|m{0.10\textwidth}|m{0.24\textwidth}|}
			\hline
			\textbf{Object} & \makecell[c]{\textbf{RCS}} & \makecell[c]{\textbf{Observation Direction}} \\
			\hline
			Sphere (radius $R$)             & \makecell[c]{$\pi {R^2}$} & Any direction  \\
			\hline
			Flat circular plate (radius $a$)
			&\makecell[c]{ $\frac{{4{\pi ^3}{a^4}}}{{{\lambda ^2}}}$} & Parallel to the normal \\
			\hline
			Cylindrical (length $L$, radius $a$) & \makecell[c]{$\frac{{2\pi a{L^2}}}{\lambda }$} &  Perpendicular to the axis of symmetry \\
			\hline
			Flat plate (area $A$)&\makecell[c]{$4\pi \frac{{{A^2}}}{{{\lambda ^2}}}$}&Parallel to the normal\\
			\hline
			Ellipsoid (semi-long axis $a$, semi-short axis $b$)&\makecell[c]{$\frac{{\pi {b^2}}}{{{a^2}}}$}&Parallel to the long or short axis\\
			\hline
			Cone&\makecell[c]{$\frac{1}{{16\pi }}{\lambda ^2}{\tan ^4}\theta $}&Parallel to the axis of symmetry\\
			\hline
		\end{tabular}
	\end{center}
	\label{11}
\end{table*}

\subsubsection{RCS of the target with simple shape}

The RCS for the target with a simple geometric shape 
can be accurately obtained, 
especially for the point target, 
whose scattering property
is isotropic, making the RCS independent of 
the angle of incidence. 
For instance, the RCS of a uniformly dense metal sphere with 
radius $R$ can be straightforwardly calculated as $\pi {R^2}$ \cite{mahafza2022radar}. 
For complicated shapes such as uniform circular plates with radius $R$, 
the scattering properties are angle-dependent, 
leading to varying RCS in different incident angles, 
which can be expressed as \cite{mahafza2022radar} 

\begin{equation}
\sigma_C=\frac{{4{\pi ^3}{R^4}}}{{{\lambda ^2}}}[\frac{{2{J_1}(\frac{{4R\sin \psi }}{\lambda })}}{{\frac{{4R\sin \psi }}{\lambda }}}]{\cos ^2}\psi, 
\end{equation} 
where $\lambda$ is the wavelength, 
$\psi$ is the angle between the incidence ray and 
the plane's normal vector, 
and ${J_1}( \cdot )$ is the first-order Bessel function.
The RCS of the target with different geometric shapes is 
listed in TABLE \ref{11} \cite{mahafza2022radar}.

\subsubsection{RCS of the target with complicated shape}

The above calculation methods of RCS are feasible 
for the target with a simple shape, 
as shown in TABLE \ref{11}. 
In the practical scenario, 
a complicated target typically includes discrete components 
with complex geometries, 
making it challenging to establish accurate RCS models.
Therefore, in the modeling methods for the RCS of the object with complex shape, the object is considered as a collection of simple scatterers. 
Due to the inflexibility and high computational complexity of the statistical averaging methods, 
various deterministic modeling methods are proposed to accurately model the RCS of complicated targets, 
such as PO method \cite{chatzigeorgiadis2004development, perez1994application}, 
GO method \cite{uluisik2008radar}, 
SBR method \cite{tao2008kd, bennani2012rcs}, 
MoM \cite{weinmann2006ray}, 
graphical electromagnetic calculation \cite{rius1993high}, etc. 
Several representative deterministic modeling methods 
are summarized as follows. 

\begin{itemize}
	\item[$\bullet$] \textbf{PO method} \cite{chatzigeorgiadis2004development, perez1994application}: 
    The PO method calculates the scattered field across the target 
    by integrating the induced electric field on the surface of target  
    to derive the target's RCS. 
    This method relies on the following conditions:
    1) the curvature radius of target’s surface is larger than the wavelength; 
    2) the target satisfies the far-field condition, 
    which necessitates detecting the target from a distance ten times 
    greater than the wavelength. 
    For example, in terms of millimeter-wave radar, 
    the PO method can quickly and accurately 
    analyze a target's scattered field 
    as long as the distance between the target and the radar 
    is greater than 10 \text{cm}. 
    However, 
    the PO method neglects the impact of the discontinuities of target surface on the induced current, which leads to the inaccuracy of the RCS calculation results.
\end{itemize}

\begin{itemize}
	\item[$\bullet$] \textbf{GO method} \cite{uluisik2008radar}: 
    The GO method involves dividing the echo signals 
    from a target into illuminated regions 
    and shadowed regions and then determines the scattering centers 
    using Fermat's principle. 
    In this method, the scattered field of the target 
    is calculated as the summation of scattered fields 
    reflected from all scattering centers, 
    enabling the calculation of the target's RCS. 
    This method has stringent requirements:
    1) the target's size must be significantly 
    larger than the incident wavelength;
    2) the electromagnetic wave must be isotropic, 
    therefore approximate to a plane wave.	
    Radar energy propagation can be approximated by electromagnetic waves 
    propagating in tubes. And the influence between each beam is small. 
    However, the GO method does not consider the scattering effect 
    caused by diffraction. 
    When the target has no curvature or single curvature, 
    the obtained RCS will be infinite.
\end{itemize}

\begin{itemize}
	\item[$\bullet$] \textbf{Signal-based ray-tracing (SBR) method} \cite{tao2008kd, bennani2012rcs}: 
The SBR method involves emitting rays in various directions through 
the transmitter to study the reflection 
characteristics of the rays in space 
and then predicts all paths during the signal propagation process. 
Moreover, the theory of electromagnetic wave propagation is applied to 
calculate the received power and time delay spread of 
each ray to analyze the channel characteristics.
The SBR method combines the advantages of PO method and GO method, 
possessing high accuracy in RCS modeling.
Therefore, it is widely adopted in direct modeling of RCS. 
However, the SBR method has high computational 
complexity\cite{smit2012comparison}. 
\end{itemize}

\begin{itemize}
	\item[$\bullet$] \textbf{Uniform theory of diffraction (UTD) method} \cite{weinmann2009utd}: 
The UTD method effectively combines the advantages of PO and GO methods. 
It divides the scattering into physical optical, 
geometric optical, and diffraction regions, 
allowing for separated RCS calculations in each region. 
UTD is particularly suitable for accurately estimating RCS 
for large and intricate targets, 
as it takes into account the factors 
such as multiple reflections and target diffractions. 
However, it is associated with high computational complexity and 
does not consider the impact of target surface roughness on RCS. 
As a result, it can not provide accurate RCS calculations for the scattering scenarios with close-range. 
Moreover, Weinmann proposed a hybrid GO-PO-UTD method 
to enhance the estimation accuracy of target RCS \cite{weinmann2009utd}.  
\end{itemize}

\begin{itemize}
	\item[$\bullet$] \textbf{Method of moment (MoM)} \cite{perotoni2007study}:
The MoM method regards the radar signal as a vector 
consisting of the target scattering center and scattering intensity. 
Then, the matrix decomposition methods are utilized to 
find the RCS of target.
The MoM method is suitable for the target with uniform scattering 
distribution and appropriate size, 
and is particularly effective in calculating 
the low-frequency scattering of target.
Due to the robustness and noise resistance capability, 
the MoM method is preferred in practice. 
However, the MoM method has drawbacks such as 
prolonged computation times and sensitivity to parameter selection. 
To tackle these problems, 
an improved MoM method is introduced 
to enhance the accuracy of RCS \cite{perotoni2007study}.
\end{itemize}

\begin{itemize}
	\item[$\bullet$] \textbf{Fast multipole method (FMM)} \cite{coifman1993fast, nie2019hybrid}:
The FMM method utilizes a hierarchical network architecture to 
decompose the target into small components. 
Then, a fast multi-pole algorithm is used to 
calculate the scattered field of target, 
thus obtaining the RCS of target \cite{coifman1993fast}.    
The FMM method is effective in accurately determining the RCS 
for various targets. 
As the complexity of target increases, 
the computation time increases correspondingly. 
However, there is still an advantage over the MoM algorithm 
in terms of computational complexity \cite{nie2019hybrid}.
\end{itemize}

\begin{itemize}
	\item[$\bullet$] \textbf{Finite-difference time-domain (FDTD) method} \cite{moss2002finite}:
The FDTD method calculates the RCS of target by discretizing 
in both space and time domains
to solve Maxwell's equations,
resulting in the spatio-temporal distribution 
of the electromagnetic field \cite{moss2002finite}. 
The FDTD method stands out in accurately computing the RCS of 
complex-shaped target 
by effectively simulating its reflection and scattering 
characteristics. 	
However, the FDTD method also faces the challenges such as 
high computational complexity,
intricate parameter selection, 
and specific medium property requirements. 
Therefore, practical applications demand 
parameter optimization and 
low complexity algorithm. 
\end{itemize}    

\subsection{Statistical Modeling of Target RCS}

\begin{table*}[htbp]
\caption{Characteristics of different statistical models of RCS}
\renewcommand{\arraystretch}{1.7}
\begin{center}
	\begin{tabular}{|m{0.2\textwidth}<{\centering}|m{0.10\textwidth}|m{0.2\textwidth}|m{0.10\textwidth}|m{0.15\textwidth}|}
		\hline
		\textbf{Type} & \makecell[c]{\textbf{Parameters}} & \makecell[c]{\textbf{Computational complexity}}&\makecell[c]{\textbf{Accuracy}}&\makecell[c]{\textbf{References}} \\
		\hline
		Swerling I & \makecell[c]{Single } &  \makecell[c]{Low}& \makecell[c]{Low}& \makecell[c]{\cite{swerling1997radar,swerling1970recent}}  \\
		\hline
		Swerling II & \makecell[c]{Single } &  \makecell[c]{Low}& \makecell[c]{Low}& \makecell[c]{\cite{swerling1997radar,swerling1970recent}}  \\
		\hline
		Swerling III & \makecell[c]{Double} &  \makecell[c]{Low}& \makecell[c]{Low}& \makecell[c]{\cite{swerling1997radar,swerling1970recent}}  \\
		\hline
		Swerling IV & \makecell[c]{Double} &  \makecell[c]{Low}& \makecell[c]{Low}& \makecell[c]{\cite{swerling1997radar,swerling1970recent}}  \\
		\hline
		Swerling V & \makecell[c]{Single } &  \makecell[c]{Low}& \makecell[c]{Low}& \makecell[c]{\cite{swerling1997radar,swerling1970recent}}  \\
		\hline
		Chi-square distribution & \makecell[c]{Double} &  \makecell[c]{Low}& \makecell[c]{Low}& \makecell[c]{\cite{meyer1973radar,xu1997new,follin1984statistics,hejduk2009improved}}  \\
		\hline
		Weibull distribution & \makecell[c]{Double} &  \makecell[c]{Low}& \makecell[c]{Low}& \makecell[c]{\cite{meyer1973radar,hughes2017piecewise}}  \\
		\hline
		Log-normal distribution & \makecell[c]{Double} &  \makecell[c]{Low}& \makecell[c]{Medium}& \makecell[c]{\cite{meyer1973radar,xu1997new,follin1984statistics,liu2017influence}}  \\
		\hline
		Rice distribution & \makecell[c]{Double} &  \makecell[c]{Medium}& \makecell[c]{Low}& \makecell[c]{\cite{schiavulli2014reconstruction}}  \\
		\hline
		GMDM & \makecell[c]{Semi} &  \makecell[c]{High}& \makecell[c]{High}& \makecell[c]{\cite{wang2021developing,wang2021novel,zhuangdynamic}}  \\
		\hline
		Lejeune polynomial & \makecell[c]{None} &  \makecell[c]{High}& \makecell[c]{High}& \makecell[c]{\cite{xu1997new,barbary2015optimisation,freundorfer2015radar}}  \\
		\hline
	\end{tabular}
\end{center}
\label{111}
\end{table*}

In practical scenario, the moving target with 
constantly changing position and angle of view
will lead to the fluctuations in their cross-sectional areas, 
requiring the application of statistical methods 
to model the RCS of target \cite{borden1983statistical}. 
Statistical methods model the RCS 
within a resolution cell as a random variable 
with specific probability density function (PDF).
The PDF describes the statistical properties of  RCS, 
where it is assumed that 
the target consists of several independent scatterers 
at random positions. 
The RCS of each scatterer is independent and constant. 
The characteristics of different statistical models 
are provided in Table \ref{111} and the details are summarized 
as follows.  

\subsubsection{Swerling I-V models}

Swerling \emph{et al.} \cite{swerling1997radar, swerling1970recent} 
introduced five Swerling models (i.e. Swerling I-V), 
which are appropriate for the scenarios with fast fluctuation, slow fluctuation, and no fluctuation \cite{swerling1997radar, swerling1970recent}. 

\begin{itemize}
		\item[$\bullet$] \textbf{Swerling I model} 
  is applicable for slow fluctuating target, 
  whose RCS follows Rayleigh distribution.
	\end{itemize}
	\begin{itemize}
		\item[$\bullet$] \textbf{Swerling II model} is applicable for fast fluctuating target, whose RCS follows Rayleigh distribution.
	\end{itemize}
	\begin{itemize}
		\item[$\bullet$] \textbf{Swerling III model} is applicable for slow fluctuating target, whose RCS follows Rice distribution.
	\end{itemize}
	\begin{itemize}
		\item[$\bullet$] \textbf{Swerling IV model} is applicable for fast fluctuating target, whose RCS follows Rice distribution.
	\end{itemize}
	\begin{itemize}
		\item[$\bullet$] \textbf{Swerling V model} is applicable for non-fluctuating target, whose RCS is a constant.
\end{itemize}

In these models, 
a slowly moving target implies that the
echo signals within the same scanning period are correlated, 
and uncorrelated for different scanning periods. 
For a fast-fluctuating target, the echo signals are 
independent between pulses in the same scan. 
The Swerling I-IV models only consider fast and slow fluctuations of 
the amplitude of echo signals,
which could model a majority of  the target RCS in practice. 

The RCS of target is denoted by $\sigma$. 
Suppose that $\bar \sigma$ represents the average target fluctuation
throughout the entire observation process. 
The PDFs of Swerling I and II models are the same, which are given by\cite{swerling1997radar} 
\begin{equation}
   f(\sigma ) = \frac{1}{{\bar \sigma }}\exp \left( {\frac{\sigma }{{\bar \sigma }}} \right), \sigma  \ge 0.
\end{equation}
The PDFs of Swerling III and IV models are the same, given by 
\begin{equation}
    f(\sigma ) = \frac{{4\sigma }}{{{{\bar \sigma }^2}}}\exp \left( { - \frac{{2\sigma }}{{\bar \sigma }}} \right), \sigma  \ge 0.
\end{equation} 

\subsubsection{Chi-square distribution model \cite{xu1997new,follin1984statistics,hejduk2009improved}}

\begin{figure}[!htbp]
\includegraphics[scale=0.6]{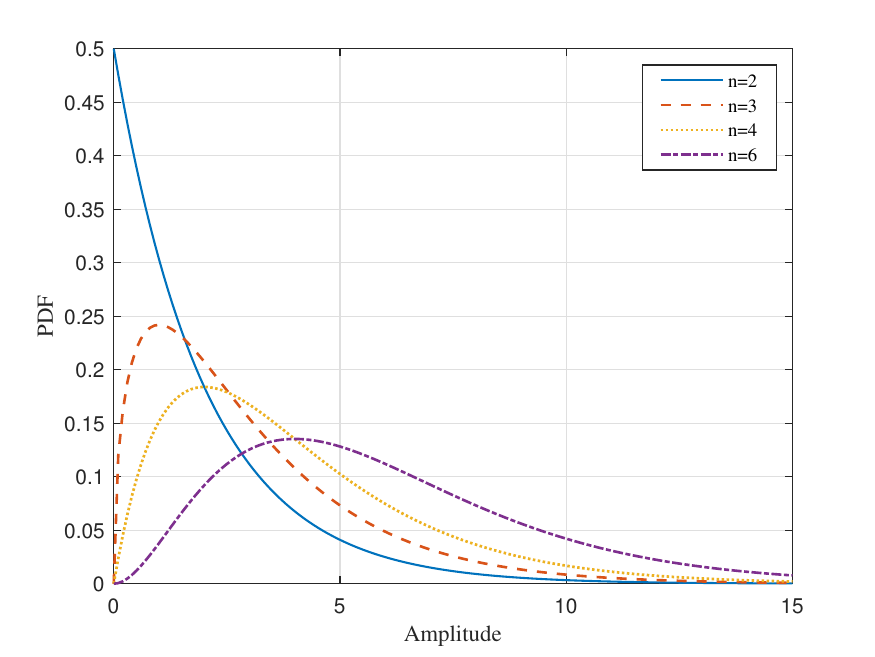}
\caption{The PDF of chi-square distribution.}
\label{chi} 
\end{figure}

Supposing that the target consists of a strong scatterer and 
a group of weak scatterers, the RCS can be modeled by chi-square distribution, which is given by\cite{hejduk2009improved}
\begin{equation}
f(\sigma ) = \frac{1}{{{2^{\frac{n}{2}}}\Gamma (\frac{n}{2})}}{\sigma ^{\frac{n}{2} - 1}}\exp ( - \frac{n}{2}),\sigma  > 0
\label{fl},
\end{equation}
where $n$ is the degree of freedom (DoF). 
Fig. \ref{chi} shows the PDF of chi-square distribution with different $n$. 
The Swerling I-V models mentioned above correspond to 
chi-square distributions with DoF of $1$, $2$, $N$, $2N$ or infinity. 
The chi-square distribution model is more applicable when the 
detection target is a regularly shaped object.

\subsubsection{Weibull distribution model \cite{hughes2017piecewise}}
\begin{figure}[!htbp]
		\includegraphics[scale=0.6]{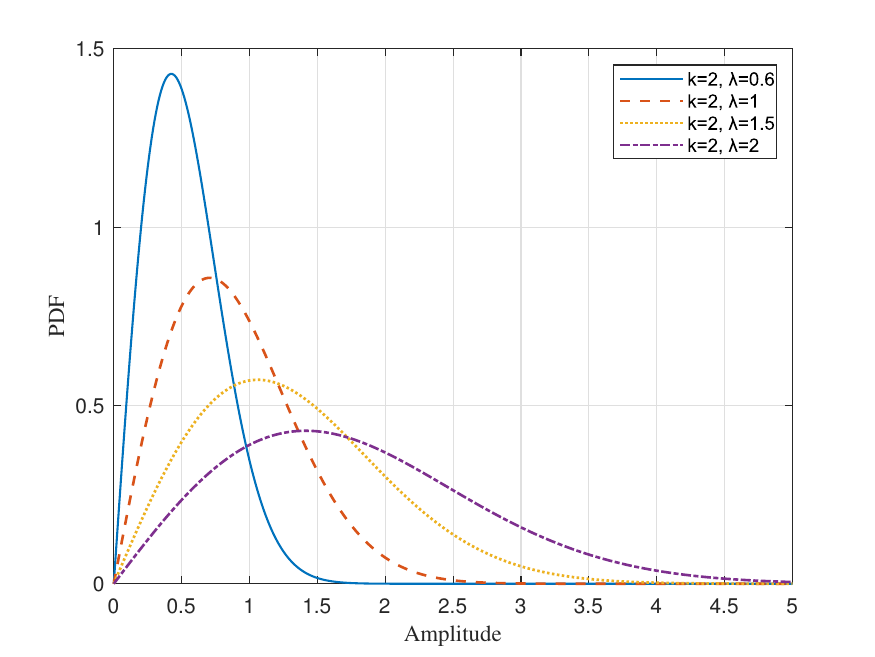}
		\caption{The PDF of Weibull distribution with different $\lambda$.}
		\label{weibull1}
	\end{figure}

\begin{figure}[!htbp]
		\includegraphics[scale=0.6]{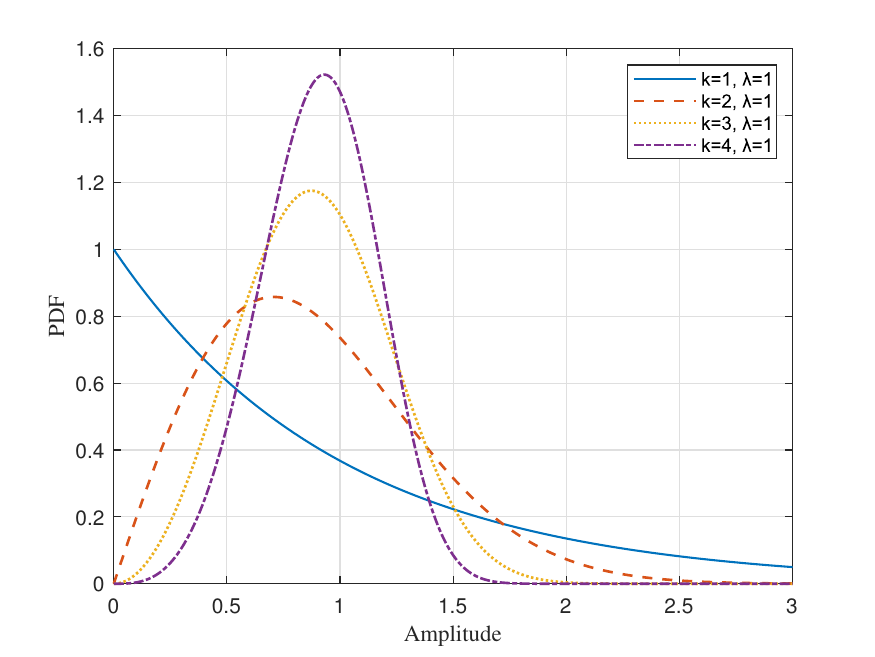}
		\caption{The PDF of Weibull distribution with different $k$.}
		\label{weibull2}
\end{figure}

The Weibull distribution is commonly utilized to model the RCS of moving target, which is given by\cite{hughes2017piecewise}
\begin{equation}
f(\sigma; k, \lambda ) = \frac{k}{\lambda }{(\frac{\sigma }{\lambda })^{k - 1}}\exp ( - \frac{k}{\lambda }),\sigma  > 0,\lambda  > 0,k > 0,
\end{equation}
where $k$ is the shape factor, $\lambda$ is the scale factor.
Fig. \ref{weibull1} shows the PDF of Weibull distribution with 
different $\lambda$ and 
Fig. \ref{weibull2} shows the PDF of 
Weibull distribution with different $k$. 
It is worth mentioning that the Weibull model approximately fits the chi-square distribution when the shape and scale factors are known. 


\subsubsection{Log-normal model \cite{xu1997new,follin1984statistics}}

\begin{figure}[!htbp]
		\includegraphics[scale=0.6]{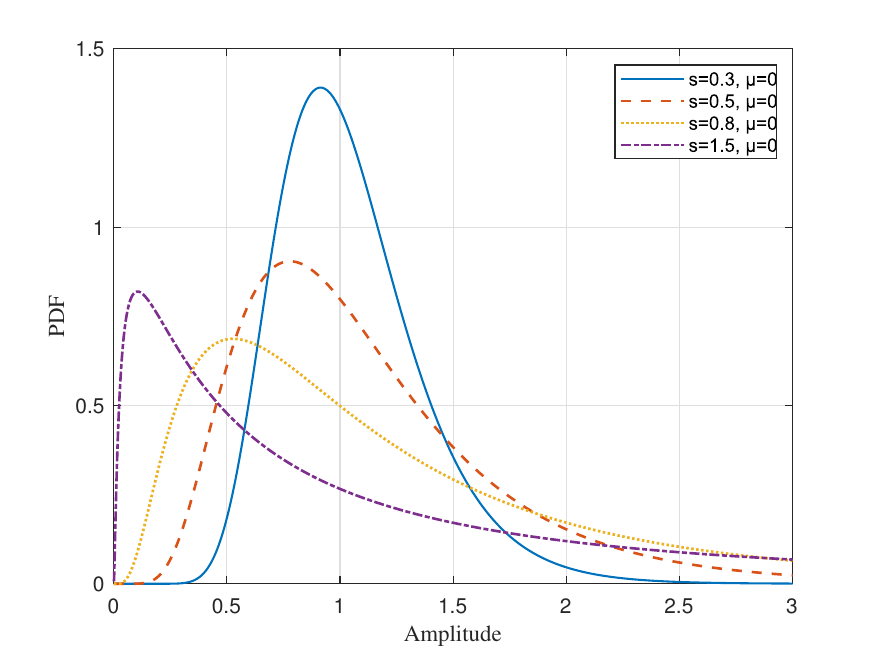}
		\caption{The PDF of log-normal distribution with different $s$.}
		\label{ln1}
	\end{figure}

\begin{figure}[!htbp]
		\includegraphics[scale=0.6]{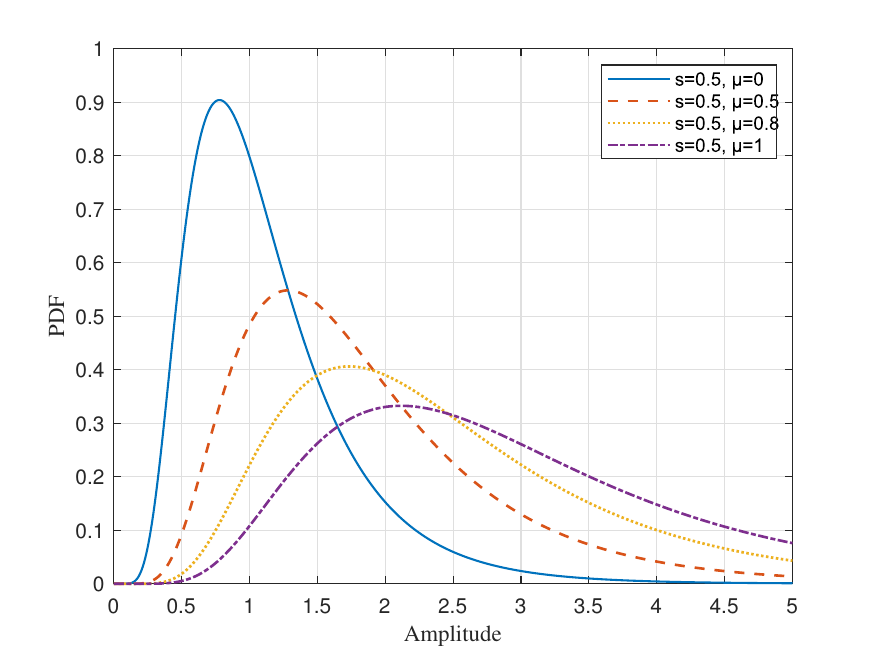}
		\caption{The PDF of log-normal distribution with different $\mu $.}
		\label{ln2}
\end{figure}

Wang \emph{et al.} \cite{wang2015analysis} discovered that when the 
RCS distribution is a single-peaked distribution, 
the log-normal distribution model has higher accuracy 
compared with the Weibull and 
chi-square distributions. 
The PDF of log-normal distribution can be expressed by \cite{ward2010use}
\begin{equation} 
f(\sigma ;\mu ,s) = \frac{1}{{s\sigma \sqrt {2\pi } }}\exp ( - {(\ln \sigma  - \mu )^2}),\sigma  > 0,
\end{equation}
where $s$ is the standard deviation and $\mu$ is the variance.
Fig. \ref{ln1} and Fig. \ref{ln2} show the PDF of log-normal distribution with 
different $s$ and $\mu$, respectively. 
It is observed that the mean affects the position of the curve, 
with an increase or decrease in the mean directly causing 
the center axis of the curve to move left or right.
Beside, the variance influences the width and height of the curve, 
with an increase in variance causing the curve to become wider and 
the tails flatter.
The log-normal distribution is effective for modeling 
the RCS of the target with large size and high velocity. 
However, its accuracy is diminishing with the increase of the target's
velocity \cite{liu2017influence}.

	
\subsubsection{Rice distribution model \cite{schiavulli2014reconstruction}}
\begin{figure}[!htbp]
		\includegraphics[scale=0.6]{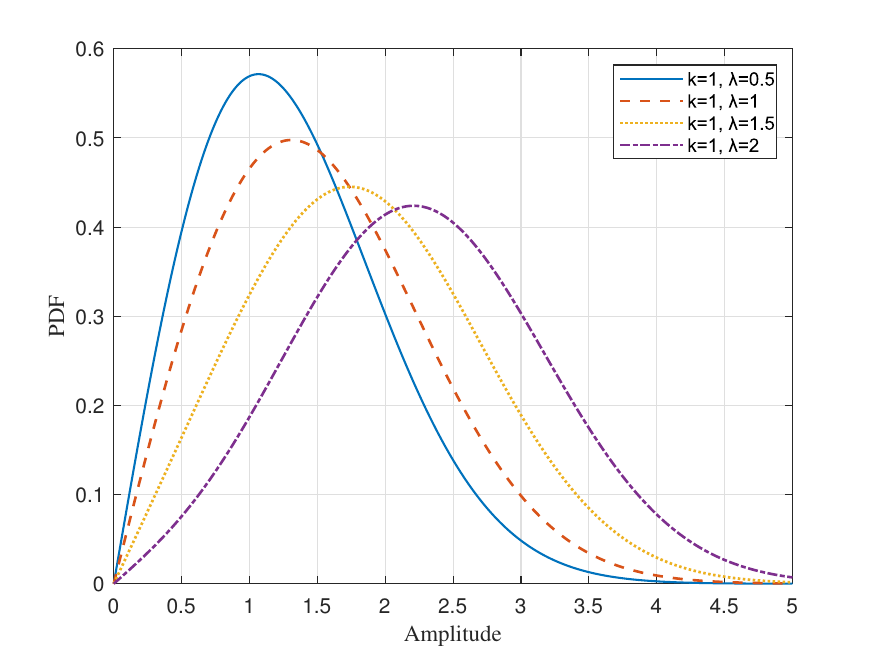}
		\caption{The PDF of Rice distribution with different $\lambda$.}
		\label{rice1}
	\end{figure}

\begin{figure}[!htbp]
		\includegraphics[scale=0.6]{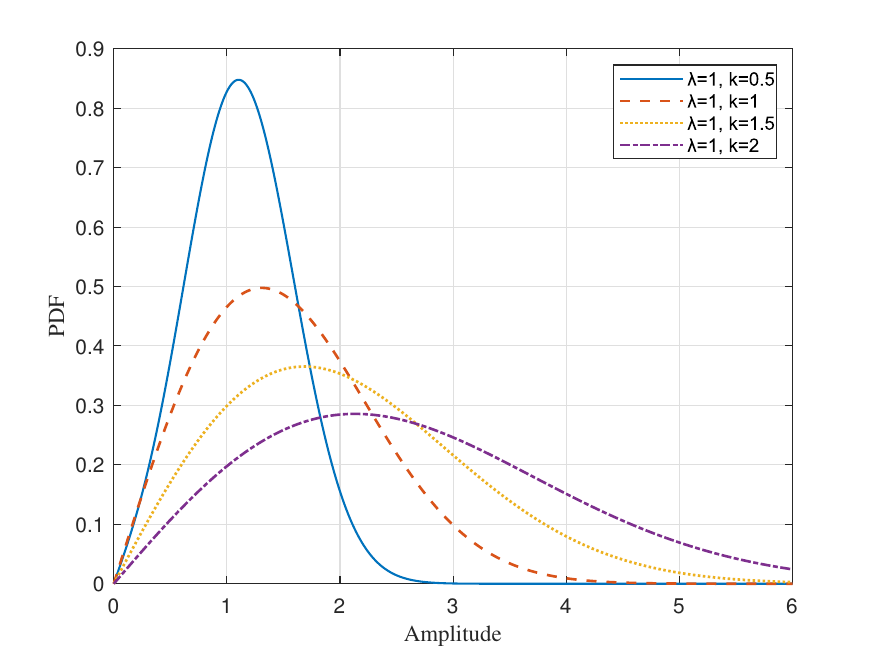}
		\caption{The PDF of Rice distribution with different $k$.}
		\label{rice2}
\end{figure}

The PDF of the Rice distribution is\cite{talukdar1991estimation} 
\begin{equation}
		f(\sigma;\lambda ,k) = \frac{1}{k}\exp \{  - \lambda  - \frac{\sigma }{k}\} {I_0}(2\sqrt {\frac{{\lambda \sigma }}{k}} ),\sigma  > 0,
\end{equation}
where ${I_0}( \cdot )$ is the $0$-order Bessel function, 
$\lambda$ is the scale factor, 
and $k$ is the shape factor. 
Fig. \ref{rice1} and Fig. \ref{rice2} show the PDF of Rice distribution with 
different $\lambda$ and $k$, respectively. 
Moreover, the Rice distribution is similar to the 
chi-square distribution, 
particularly in the case that 
the values of $\lambda$ and $k$ are 
approximately equal. 
However, in practical scenarios, 
the RCS of target is dynamic, 
and the Rice distribution model demonstrates poor performance 
due to the diverse factors influencing target characteristics.


\subsubsection{Gaussian mixture density mode (GMDM) \cite{wang2021developing, wang2021novel, zhuangdynamic}}

When the movement and trajectory of target are changing,
the above RCS statistical models are 
no longer available \cite{li2022rcs}. 
Zhuang \emph{et al}. \cite{zhuang2014statistical} proposed a 
Gaussian mixture density mode (GMDM)-based statistical model 
to describe the RCS of the target with time-varying mobility. 
GMDM is a semi-parametric model that utilizes a mixture of 
multiple Gaussian distributions. 
Then, the PDF of the $M$-order GMDM can be expressed as \cite{wang2021developing,wang2021novel}
\begin{equation}
f(\sigma ;Q) = \sum\limits_{i = 1}^M {\frac{{{a_i}}}{{\sqrt {2\pi \sigma _i^2} }}} \exp ( - \frac{{{{(\sigma  - {\mu _i})}^2}}}{{2{\sigma _i}}}),
\end{equation}
where ${\mu _i}$ and $\sigma _i^2$ are the mean and variance of the $i$-th Gaussian distribution, respectively, $a_i$ 
satisfies $\sum\limits_{i = 1}^M {{a_i}}  = 1$, and $Q$ denotes 
$\{ {a_1},{a_2}, \cdots {a_M};{\mu _1},{\mu _2}, \cdots {\mu _M};\sigma _1^2,\sigma _2^2, \cdots ,\sigma _M^2\}$.
Zhuang \emph{et al}. \cite{zhuangdynamic} applied the GMDM to model the dynamic RCS of aircraft, whose accuracy improves with the increase of $M$. However, it concurrently results in high complexity.

\subsubsection{Legendre orthogonal polynomials (LOP) Model 
\cite{xu1997new,barbary2015optimisation,freundorfer2015radar}}

The parametric model establishes a certain PDF 
with finite parameters. 
In contrast, non-parametric model is more flexible than 
parametric model. 
In order to obtain accurate RCS model of 
the target with high mobility, 
the non-parametric model of RCS is widely studied. 
In particular, LOP model is a common non-parametric model, 
which requires a combination of central moments and LOP to
approximate the PDF of the non-parametric model \cite{barbary2015optimisation}. 
Hence, it is only necessary to calculate the central moments 
of all $N$-th order observed values 
and the LOP will fit the RCS with arbitrary accuracy. Then, the PDF of LOP model is given by \cite{freundorfer2015radar}
\begin{equation}
		{f_{}}(\sigma ) = \frac{1}{{{\sigma _L}}}f(\frac{{\sigma  - \bar \sigma }}{{{\sigma _L}}}) = \frac{1}{{{\sigma _L}}}\sum\limits_{i = 0}^\infty  {{a_i}} {L_i}(\frac{{\sigma  - \bar \sigma }}{{{\sigma _L}}}),
\end{equation}
where ${\sigma _L}$ is the interval length, 
i.e. ${\sigma _L} = {\sigma _{\max }} - {\sigma _{\min }}$, ${L_i}( \cdot )$ is Legendre polynomial, 
and ${a_i}$ is determined by the $k$-th 
central moment of $\sigma$. 
The RCS fitting accuracy is improving when the order $N$ is increased from 15 to 25. 
The non-parametric model fits well in practice with the cost of high complexity.
Implementing the Lejeune polynomial model still 
lacks fast algorithms.

\section{Clutter RCS Modeling}
\label{Sec_Clutter_RCS}

Radar clutter refers to the scattered echos caused by 
the objects other than the target, 
characterized by clutter RCS\cite{knott2004radar}. 
Clutter RCS is influenced by wavelength, polarization, 
and angle of incidence. 
The clutter RCS is a crucial performance metric 
describing the scattering 
cross-sectional area of clutter, 
offering valuable insights into the performance analysis and 
signal processing in radar sensing. 
This section introduces the classification of clutter, 
followed by an in-depth exploration on the various statistical models 
of clutter RCS. 

\subsection{Classification of Clutter}

According to the application scenarios, 
clutter is categorized into three types, 
namely sea clutter, ground clutter, and meteorological clutter 
\cite{barton1988modern,levanon1988radar}.
Sea clutter is the scattered echoes produced by incident electromagnetic 
waves passing over the sea surface \cite{haykin2002uncovering}. 
The scattering coefficient of sea clutter 
is related to the wind speed, 
wind direction, angle of incidence, 
and the state of sea surface 
\cite{xin2016deterministic,simon2013nonlinear,liu2017analysis}.
Ground clutter is defined as the extraneous echoes received by 
radar when detecting the target on ground. 
Ground clutter is impacted by the angle of incidence, 
the ground's physical characteristics, 
and the coverage area of radar beam \cite{louf2019integrated}.  
Meteorological clutter is the scattered echoes 
when a radar signal passes through clouds, rain, snow, hail, etc. 
Meteorological clutter is mainly related to 
the state of air and the coverage volume of radar beam.

\subsection{Statistical Modeling of Clutter RCS}

Radar clutter is the sum of numerous echoes 
reflected from scatterers within the coverage of radar. 
The scatterers will cause variations of echoes, 
resulting in random fluctuations in clutter RCS \cite{kumlu2019clutter,zebiri2021radar}. 
The RCS characteristic of clutter is usually described by 
the PDF of clutter. 
Moreover, the commonly used statistical models including 
Rayleigh distribution \cite{hou1989detection}, 
log-normal distribution \cite{sayama2021detection,kim2022accurate}, 
Weibull distribution \cite{garcia2020square}, 
and $K$-distribution \cite{medeiros2023cfar,zhao2019robust}. 
Table \ref{222} summarizes the characteristics of 
statistical models 
of clutter RCS. 
The details of these distributions are introduced as follows.

\begin{table*}[htbp]
	\caption{Characteristics of statistical models of clutter RCS}
	\renewcommand{\arraystretch}{1.7}
	\begin{center}
		\begin{tabular}{|m{0.2\textwidth}<{\centering}|m{0.10\textwidth}|m{0.2\textwidth}|m{0.10\textwidth}|m{0.15\textwidth}|}
			\hline
			\textbf{Category} & \makecell[c]{\textbf{Parameter}} & \makecell[c]{\textbf{computational complexity}}&\makecell[c]{\textbf{Accuracy}}&\makecell[c]{\textbf{References}} \\
			\hline
			Rayleigh Distribution & \makecell[c]{Double } &  \makecell[c]{Low}& \makecell[c]{Low}& \makecell[c]{\cite{hou1989detection}}  \\
			\hline
			Log-normal distribution & \makecell[c]{Double } &  \makecell[c]{Low}& \makecell[c]{Medium }& \makecell[c]{\cite{sayama2013suppression,farina1997high}}  \\
			\hline
			Weibull distribution & \makecell[c]{Double} &  \makecell[c]{Low}& \makecell[c]{Medium }& \makecell[c]{\cite{sekine1981weibull,garcia2020square,sayama2001weibull}}  \\
			\hline
			$K$-distribution & \makecell[c]{Double} &  \makecell[c]{Medium}& \makecell[c]{High}& \makecell[c]{\cite{hou1989detection,farina1997high}}  \\
			\hline
			Generalized composite clutter model & \makecell[c]{Three } &  \makecell[c]{High}& \makecell[c]{High}& \makecell[c]{\cite{qin2012cfar,gao2016scheme,qin2012statistical,huang2021statistical,shuang2006parameters,ollila2012compound}}  \\
			\hline			
		\end{tabular}
	\end{center}
	\label{222}
\end{table*}

\subsubsection{Rayleigh distribution clutter model}

Supposing that the clutter RCS is ${\sigma _C}$ and 
the variance is ${k^2}$, 
the PDF of the Rayleigh distribution clutter model is given by\cite{blasch2004fusion}
\begin{equation}
f({\sigma _C}) = \frac{{2{\sigma _C}}}{{{k^2}}}\exp ( - \frac{{{\sigma _C}^2}}{{{k^2}}}).
\end{equation}

\begin{figure}[!htbp]
		\includegraphics[scale=0.6]{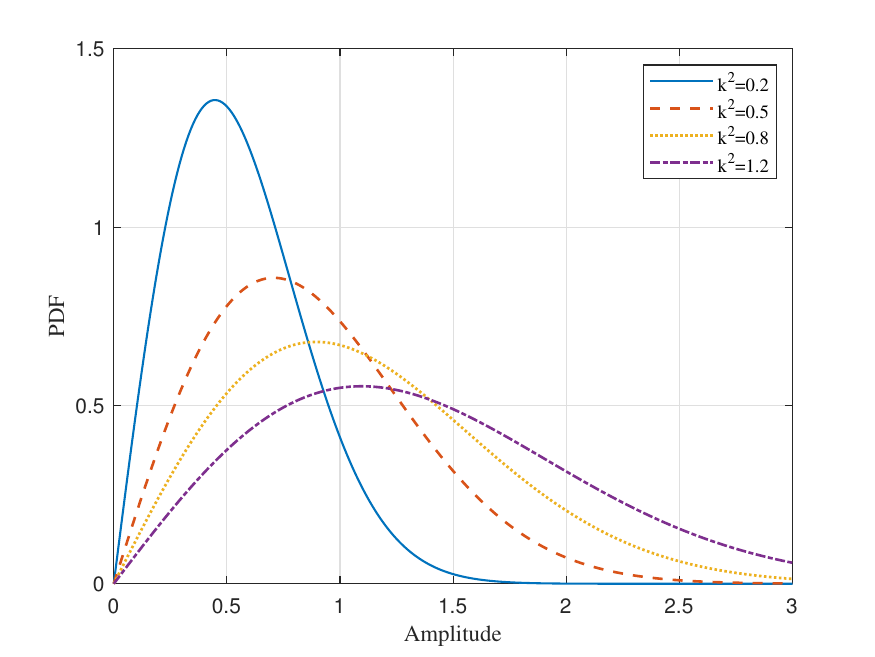}
		\caption{The PDF of Rayleigh distribution with different ${k^2}$.}
		\label{ruili}
\end{figure}

When the clutter cell consists of several independent weak scatterers, 
the RCS of clutter can be modeled by Rayleigh distribution, 
which is suitable for describing meteorological clutter and ground clutter.
Fig. \ref{ruili} shows the PDF of Rayleigh distribution 
with different variances.
It can be observed that when ${k^2}$ equals 1, the Rayleigh distribution is equivalent to the chi-square distribution with 2 degrees of freedom. As the ${k^2}$ increases, the PDF curve becomes flatter.

\subsubsection{Log-normal distribution clutter model}

The PDF of log-normal distribution can be expressed as \cite{xu1997new}
\begin{equation}
f(\sigma_C ;\mu ,s) = \frac{1}{{s\sigma_C \sqrt {2\pi } }}\exp ( - {(\ln \sigma_C  - \mu )^2}),\sigma  > 0,
\end{equation}
where $\ln \sigma _C^{}$ follows normal distribution 
with standard deviation $s$ and variance $\mu$. 
The log-normal distribution is suitable for 
modeling the RCS of the ground clutter with small 
incidence angles and simple topography, 
and sea surface clutter with high-resolution requirements for radar systems. 
The shape parameter of log-normal distribution usually takes values between  
$0.5$ and $1.3$ \cite{anastassopoulos1994generalized}. 
Fig. \ref{ln1} and Fig. \ref{ln2} 
show the PDFs of log-normal distribution 
with different $s$ and $\mu $, respectively. 
It is revealed that when the $\mu $ is unchanged, 
the larger the $s$, 
the smaller the PDF peak and the longer its tail.

\subsubsection{Weibull distribution clutter model }
The PDF of the Weibull distribution model is \cite{shi2013radar}
\begin{equation}
f(\sigma_C ;k,\lambda ) = \frac{k}{\lambda }{(\frac{\sigma_C }{\lambda })^{k - 1}}\exp ( - \frac{k}{\lambda }),\sigma_C  > 0,\lambda  > 0,k > 0,
\end{equation}
where $k$ is the shape factor, 
and $\lambda$ is the scale factor.
Fig. \ref{weibull1} and Fig. \ref{weibull2} show 
the PDF of the Weibull distribution 
with different $\lambda$ and $k$, respectively.
Note that when the shape parameter $k = 2$, 
the Weibull distribution is equivalent to the Rayleigh distribution, 
characterized by its heavy tail. 
It is commonly applied in the scenarios
with high-resolution radar systems 
and shallow incidence angles\cite{manavalan2018review}.

\subsubsection{$K$-distribution clutter model}
The PDF of $K$-distribution model can be expressed as\cite{watts1985radar}
\begin{equation}
		f({\sigma _C};k,\lambda ) = \frac{2}{{k\Gamma (\lambda  + 1)}}{(\frac{{{\sigma _C}}}{{2k}})^{\lambda  + 1}}{I_\lambda }(\frac{{{\sigma _C}}}{k}),{\sigma _C} > 0,\lambda  >  - 1,
\end{equation}
where $k$ is the shape factor, 
$\lambda$ is the scale factor, 
${I_\lambda }( \cdot )$ is the $\lambda$-order 
modified Bessel function of the first kind 
and $\Gamma (\cdot)$ is the gamma function. 
The $K$-distribution is widely applied in modeling 
ground and sea clutter with high-resolution radar systems and 
shallow incidence angles \cite{jao1984amplitude}.
Rayleigh, log-normal, and Weibull distributions 
are all derived from single-point statistics, 
lacking the ability to 
model the temporal and spatial correlations of clutter. 
In contrast, the $K$-distribution effectively models 
the spatio-temporal characteristics of clutter.

\begin{figure}[!htbp]
		\includegraphics[scale=0.6]{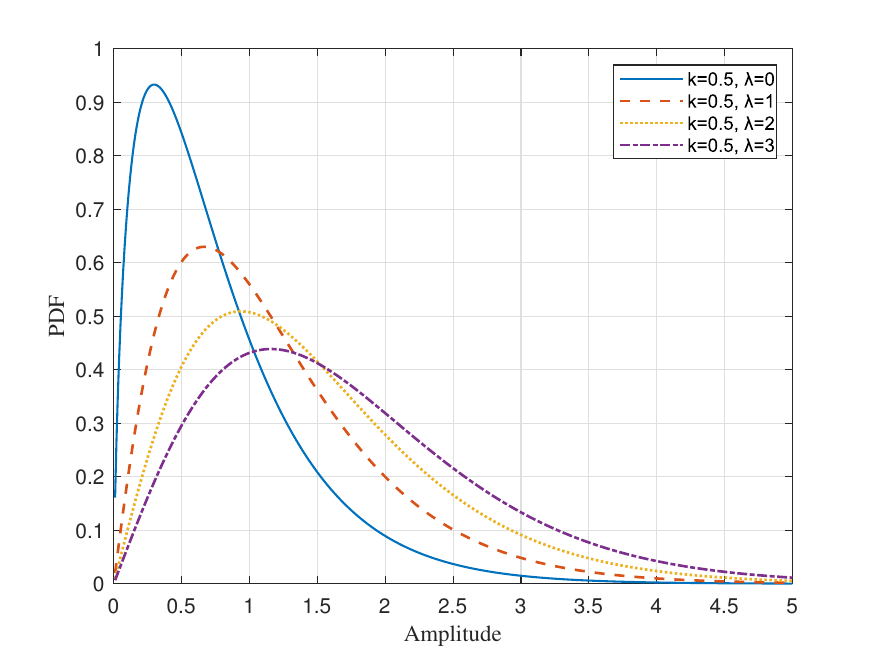}
		\caption{The PDF of $K$-distribution with different $\lambda$.}
		\label{k1}
	\end{figure}

\begin{figure}[!htbp]
		\includegraphics[scale=0.6]{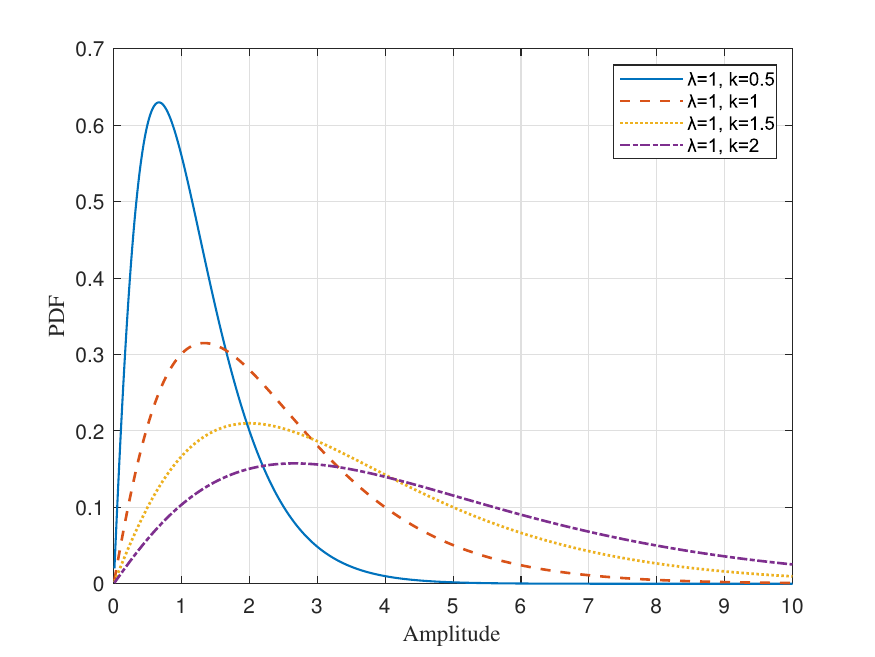}
		\caption{The PDF of $K$-distribution with different $k$.}
		\label{k2}
\end{figure}

Figs. \ref{k1} and \ref{k2} show the PDFs of $K$-distribution with different $\lambda$ and different $k$, respectively.
The $k$ parameter significantly influences the tail shape of the PDF, 
with an increase in $k$ extending the tail. 
Conversely, the $\lambda$ parameter correlates solely with 
the mean value of the clutter RCS. 
A smaller $\lambda$ results in an extended tail, 
whereas as $\lambda$ approaches infinity, 
the $k$-distribution converges towards a Rayleigh distribution.
However, the accuracy of $K$-distribution is low
for high-resolution radar systems. 
To improve the accuracy of $K$-distribution, 
$KK$-distribution is proposed to accurately simulate complex scattering effects 
in high-resolution radar systems by establishing a hybrid model of two $K$-distributions in a high-resolution radar system\cite{yanhui2006simulation}.

	
\subsubsection{Clutter composite statistical model}
With the increase in the frequency and resolution of radar systems, 
the $K$-distribution model does not satisfy 
the requirement of the modeling of 
the majority of clutters. 
Meanwhile, the composite $K$-distribution model 
describes the clutter amplitude as the product of two factors, 
namely the speckle component and the amplitude modulation component. 
In this case, the speckle component, i.e., the fast-changing component, 
is formed by the superposition of the reflected phases of a large number 
of scatterers following Rayleigh distribution.   

In high-resolution radar systems, 
a resolution cell typically contains 
a limited number of scatterers. 
The clutter exhibits non-Rayleigh characteristics 
due to the absence of phase references from numerous scatterers. 
Li \textit{et al.} \cite{li2007extended} introduced 
a generalized composite clutter RCS model 
with high flexibility,
which is especially valuable in high-resolution radar systems.
The PDF of the composite statistical model can be expressed as\cite{greco2014radar}
\begin{equation}
	f({\sigma _{C}};k,\lambda ) = \frac{{\left| b \right|}}{{\lambda \Gamma \left( k \right)}}{(\frac{{{\sigma _C}}}{\lambda })^{bk - 1}}\exp ( - {(\frac{{{\sigma _C}}}{\lambda })^b}),\lambda  > 0,k > 0,
\end{equation}
where $k$ is the shape parameter, 
$\lambda$ is the scale parameter, 
$b$ is the power parameter of 
the Generalized Compound Probability Density Function (GC-PDF).
This model is based on the premise that 
both the fast and slow variation components follow 
the generalized Gamma distribution. 
This clutter model is the exponential distribution 
when $b=1$ and ${\sigma _{C}}=1$. 
If $b=2$, ${\sigma _{C}}=\frac{1}{2}$, and 
$\lambda  =  \sigma \sqrt 2$, 
the clutter follows Gaussian distribution. 
If $b=2$ and ${\sigma _{C}}=1$, 
the clutter follows Rayleigh distribution. 
If ${\sigma _{C}}= 1$, 
the clutter follows Weibull distribution. 
Therefore, the exponential distribution, 
Gaussian distribution, Rayleigh distribution, and Weibull distribution are the special 
cases of the composite statistical model.

\section{Future Trends}

This section provides the future research trends of 
ISAC channel modeling in the era of 6G, 
including ISAC channel measurement, 
ISAC channels for new applications, 
and ISAC channel models for cooperative sensing.

\subsection{ISAC Channel Measurement}

Efficient ISAC channel modeling is critical for the 
performance evaluation of ISAC system.
Recently, the propagation characteristics 
of radar sensing channels are measured \cite{liyanaarachchi2021optimized}. 
The relation between the radar sensing channel and 
parameters such as distance and angle in an ideal 
open-field scenario is investigated.
However, the impact of non-ideal 
environment on channel modeling is rarely considered. 
In non-ideal environment, \cite{zhang2022mmwave} 
and \cite{yu2022implementation} have studied the 
relation between environmental parameters and 
ISAC channel characteristics. 
In indoor environment involving the sensing of human activities, 
\cite{tang2021multipath} studied the 
impact of wall interference 
on ISAC channels. 
However, they mainly focused on the simple 
scatterer distribution. 
Wang \emph{et al.} \cite{wang2022empirical} explored the relation between 
environmental interference and radar sensing channels.
To model the radar sensing channel, 
the traditional parameters not measured in communication channel
measurement need to be measured.
Zhang \emph{et al.} \cite{zhang2023integrated} developed 
an ISAC channel measurement platform to analyze 
the channel characteristics of RCS, 
environmental information, 
and scatterer space information, 
laying a foundation for the study of ISAC channel modeling.

\subsection{ISAC Channel Models for New Applications}

The complex electromagnetic propagation environment
in the new applications of ISAC brings 
challenge for ISAC channel modeling. 
In this section, we investigate the 
ISAC channel modeling in the new applications including 
passive sensing for target localization and tracking, 
environmental reconstruction and target imaging, 
and gesture recognition.

\subsubsection{Passive sensing for target localization and tracking} 

In target localization and tracking, 
the distance and velocity estimation  
are crucial. 
The target is commonly modeled as 
point by neglecting the volume and shape.
In this application,
ISAC channel modeling emphasizes 
the RCS characteristics of the point target 
\cite{10214383,jiang2023cooperation,jiang2023isac}. 

\subsubsection{Environmental reconstruction and target imaging} 

In this application, the accurate reconstruction of 
the positions of static, quasi-static, and dynamic scatterers 
in the surrounding environment is required. 
Meanwhile, specific scattering characteristics 
of these scatterers need to be modeled. 
In this application, ISAC channel modeling 
needs to consider various surface materials 
and features of scatterers \cite{lu2023isac}.
	
\subsubsection{Gesture recognition} 

In gesture recognition,
the analysis of micro-Doppler information on 
moving targets is critical. 
In this application, 
ISAC channel modeling requires detailed parameters of gestures 
to ensure accurate recognition of gesture \cite{zhang2022wi}.	

\subsection{ISAC Channel Model for Cooperative Sensing} 

\begin{figure}[!t]
	\centering
	\includegraphics[width=0.45\textwidth]{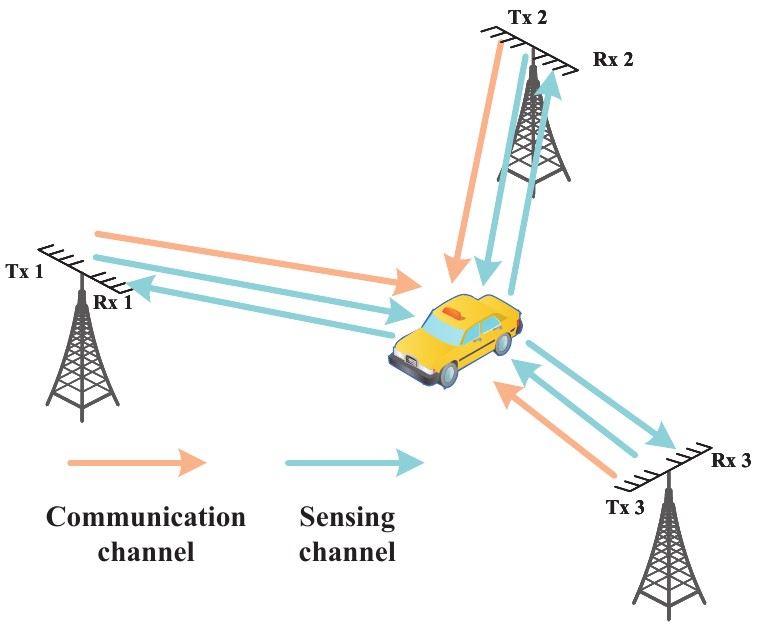}
	\caption{{The scenario of cooperative sensing.}}
	\label{ cooperative sensing}
\end{figure}

Recently, the cooperative sensing in ISAC system 
has attracted significant 
attention to break through the limitation of single-node sensing 
\cite{yang2023multi, wei2023integrated}, as shown in Fig. \ref{ cooperative sensing}.
Cooperative sensing fuses the sensing information of multiple nodes 
to realize accurate and large-scale sensing. 
Considering multiple nodes simultaneously detecting targets 
in cooperative sensing,
the antenna arrays of multiple nodes can be 
viewed as a virtual antenna array
and an equivalent radar sensing channel can be established 
based on the correlation between BSs \cite{10411109,10273396,10226276}. 

\section{Conclusion}

This article provides a comprehensive survey on ISAC channel modeling. 
Firstly, the channel modeling methods of ISAC channels 
with active and passive sensing modes are reviewed. 
Furthermore, the channel modeling methods for target RCS and 
clutter RCS are reviewed from both deterministic 
and statistical modeling perspectives. 
Finally, the future trends of ISAC channel modeling methods 
toward emerging applications in the 6G era are summarized. This article provides a guideline for channel modeling in ISAC systems.


\bibliographystyle{IEEEtran}
\bibliography{reference}
	
\ifCLASSOPTIONcaptionsoff
\newpage
\fi

\end{document}